\documentclass[lettersize,journal]{IEEEtran}
\usepackage{amsmath,amsfonts}
\usepackage{algorithmic}
\usepackage{algorithm}
\usepackage{array}
\usepackage[caption=false,font=normalsize,labelfont=sf,textfont=sf]{subfig}
\usepackage{textcomp}
\usepackage{stfloats}
\usepackage{url}
\usepackage{verbatim}
\usepackage{graphicx}
\usepackage{cite}
\usepackage{tabularx}
\usepackage{multirow}
\usepackage{color}
\usepackage{booktabs} 
\usepackage{makecell}

\begin{document}

\title{Global Spatial-Temporal Information-based Residual ConvLSTM for Video Space-Time Super-Resolution}

%\author{Congrui Fu,~\IEEEmembership{Staff,~IEEE,}
\author{Congrui Fu, Hui Yuan,~\IEEEmembership{Senior Member,~IEEE,} Shiqi Jiang, Guanghui Zhang, Liquan Shen, and Raouf Hamzaoui,~\IEEEmembership{Senior Member,~IEEE,}
	% <-this % stops a space

\thanks{Corresponding author: Hui Yuan (e-mail: huiyuan@sdu.edu.cn).}
\thanks{Congrui Fu, and Hui Yuan are with the School of Control Science and Engineering, Shandong University, Jinan, Shandong, China.}
\thanks{Shiqi Jiang is with the School of software, Shandong University, Jinan, Shandong, China.}
\thanks{Guanghui Zhang is with School of Computer Science and Technology, Shandong University, China.}
\thanks{Liquan Shen is with the School of Communication and Information Engineering, Shanghai University, Shanghai, China.}
\thanks{Raouf Hamzaoui is with the School of Engineering and Sustainable Development, De Montfort University, LE1 9BH Leicester, U.K.}

\thanks{This work was supported in part by the National Natural Science Foundation of China under Grants 62222110 and 62172259, the Taishan Scholar Project of Shandong Province (tsqn202103001), the Natural Science Foundation of Shandong Province under Grant ZR2022ZD38, the High-end Foreign Experts Recruitment Plan of Chinese Ministry of Science and Technology under Grant G2023150003L, and the OPPO Research Fund.
}
}


\maketitle

\begin{abstract}

By converting low-frame-rate, low-resolution videos into high-frame-rate, high-resolution ones, space-time video super-resolution techniques can enhance visual experiences and facilitate more efficient information dissemination. 
We propose a convolutional neural network (CNN) for space-time video super-resolution, namely GIRNet. 
To generate highly accurate features and thus improve performance, the proposed network integrates a feature-level temporal interpolation module with deformable convolutions and a global spatial-temporal information-based residual convolutional long short-term memory (convLSTM) module. 
%There are two main innovations.
In the feature-level temporal interpolation module, we leverage deformable convolution, which adapts to deformations and scale variations of objects across different scene locations. This presents a more efficient solution than conventional convolution for extracting features from moving objects. Our network effectively uses forward and backward feature information to determine inter-frame offsets, leading to the direct generation of interpolated frame features. 
%Th other is the extraction and utilization of global spatial-temporal information.
In the global spatial-temporal information-based residual convLSTM module, the first convLSTM is used to derive global spatial-temporal information from the input features, and the second convLSTM uses the previously computed global spatial-temporal information feature as its initial cell state. 
This second convLSTM adopts residual connections to preserve spatial information, thereby enhancing the output features. 
Experiments on the Vimeo90K dataset show that the proposed method outperforms state-of-the-art techniques in peak signal-to-noise-ratio (by 1.45 dB, 1.14 dB, and 0.02 dB over STARnet, TMNet, and 3DAttGAN, respectively), structural similarity index(by 0.027, 0.023, and 0.006 over STARnet, TMNet, and 3DAttGAN, respectively), and visually.

%The source code for GIRNet can be found at xxx. on the VIMEO90K, the REDS, and the UCF101 datasets
%
\end{abstract}


\section{Introduction}

\IEEEPARstart{T}{he} growing popularity of video across various fields including information dissemination, entertainment, marketing, communication, and cultural heritage, has fueled the demand for high-definition content. This has driven significant attention to video super-resolution (VSR), which aims to enhance the visual quality of videos by transforming a low-resolution video into a high-resolution video~\cite{10420512,9941493,9919402,8723517,10288391}.
VSR finds applications in various domains such as satellite imaging~\cite{10239514}, surveillance~\cite{9530280}, video compression~\cite{VC}, face recognition~\cite{face}, object recognition~\cite{9409729}, and ultra-high-definition (UHD) video processing~\cite{uhd}.

\begin{figure}[htbp]
	\centering
	\includegraphics [width=\linewidth]{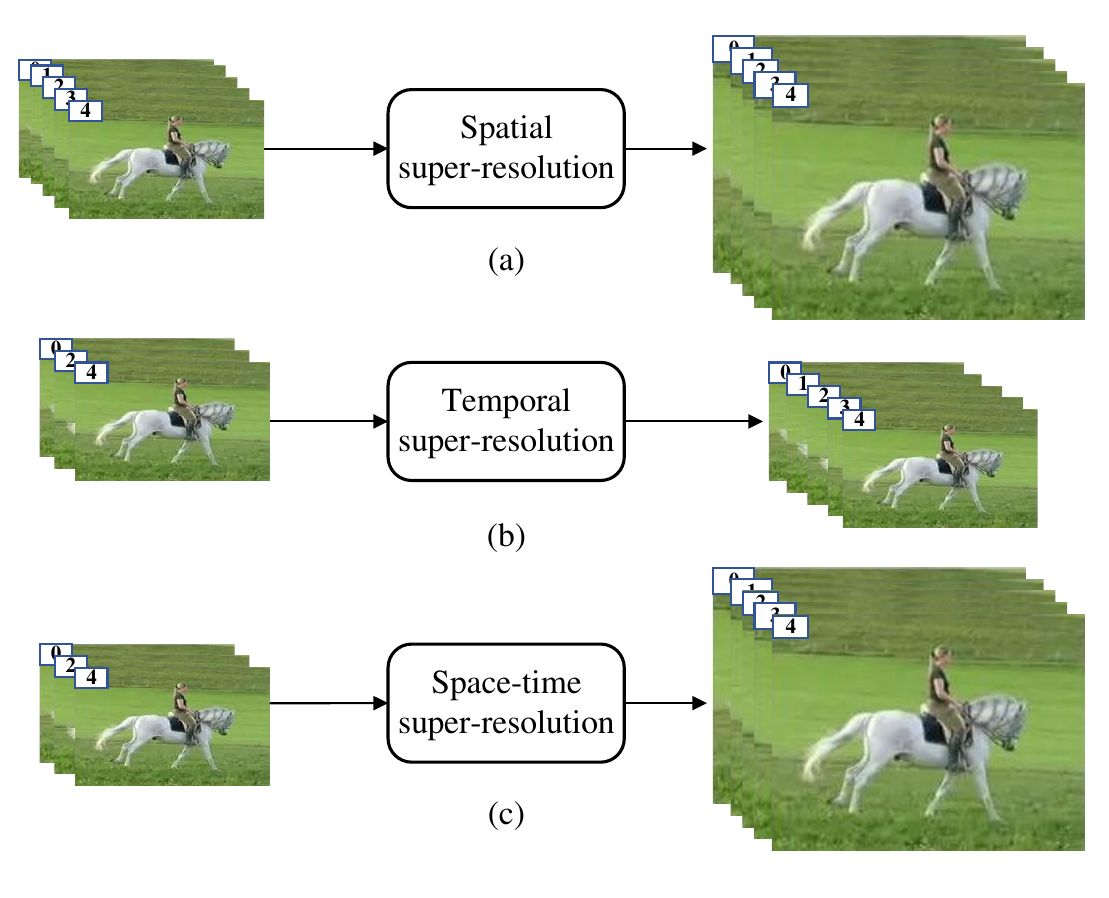}
	\caption{Video super-resolution, (a) spatial super-resolution, (b) temporal super-resolution, (c) space-time super-resolution.}
	\label{kinds}	
\end{figure}

In general, VSR refers to the process of improving the spatial resolution of each individual frame in a video (Fig.~\ref{kinds}(a)). 
Besides video spatial super-resolution (VSSR), there are two other types of VSR: video temporal super-resolution (VTSR) and video space-time super-resolution (VSTSR)~\cite{9793365}. 
In VTSR (Fig.~\ref{kinds}(b)), also known as video frame interpolation (VFI), an intermediate video frame is interpolated between existing video frames. In VSTSR, the goal is to increase both the spatial and temporal dimensions of the input video (Fig.~\ref{kinds}(c)). 
The basic mathematical model of VSR can be expressed as follows
\begin{equation}
{\textbf{Y}_{HR}} = f(\textbf{X}_{LR_1}, \textbf{X}_{LR_2},...,\textbf{X}_{LR_N}),
\end{equation}
where $\textbf{X}_{LR_1}, \textbf{X}_{LR_2},...,\textbf{X}_{LR_N}$ represent the input low-resolution (LR) video frames, and $\textbf{Y}_{HR}$ represents an output high-resolution (HR) video frame. Function $f$ denotes the super-resolution algorithm, which maps multiple low-resolution video frames into high-resolution video frames. 
In this paper, we focus on VSTSR.

One approach to achieve VSTSR is to sequentially combine VSSR and VTSR models.
However, since time and space are closely intertwined, the sequential combination may not fully leverage the inherent space-time relationship. Additionally, generating high-quality frames necessitates the utilization of state-of-the-art, computationally intensive VSSR and VFI models, resulting in unacceptable complexity.
An alternative approach is to use one-stage VSTSR methods. This approach simultaneously applies VFI and VSSR methods to low-frame-rate and low-resolution videos. 
The paramount feature of VSTSR lies in its ability to leverage the spatio-temporal information inherent in a video sequence. The efficacy of information utilization significantly impacts the model's performance. While many existing methods are built upon this premise, certain drawbacks persist. For instance, approaches using 3D convolution suffer from excessive computational demands. On the other hand, optical flow estimation methods may fall short in ensuring the precision of motion information. Consequently, there remains a need for further research aimed at enhancing the efficiency of intra-frame and inter-frame information utilization.

\begin{figure*}[!b]
	\centering
	\includegraphics [width=\linewidth]{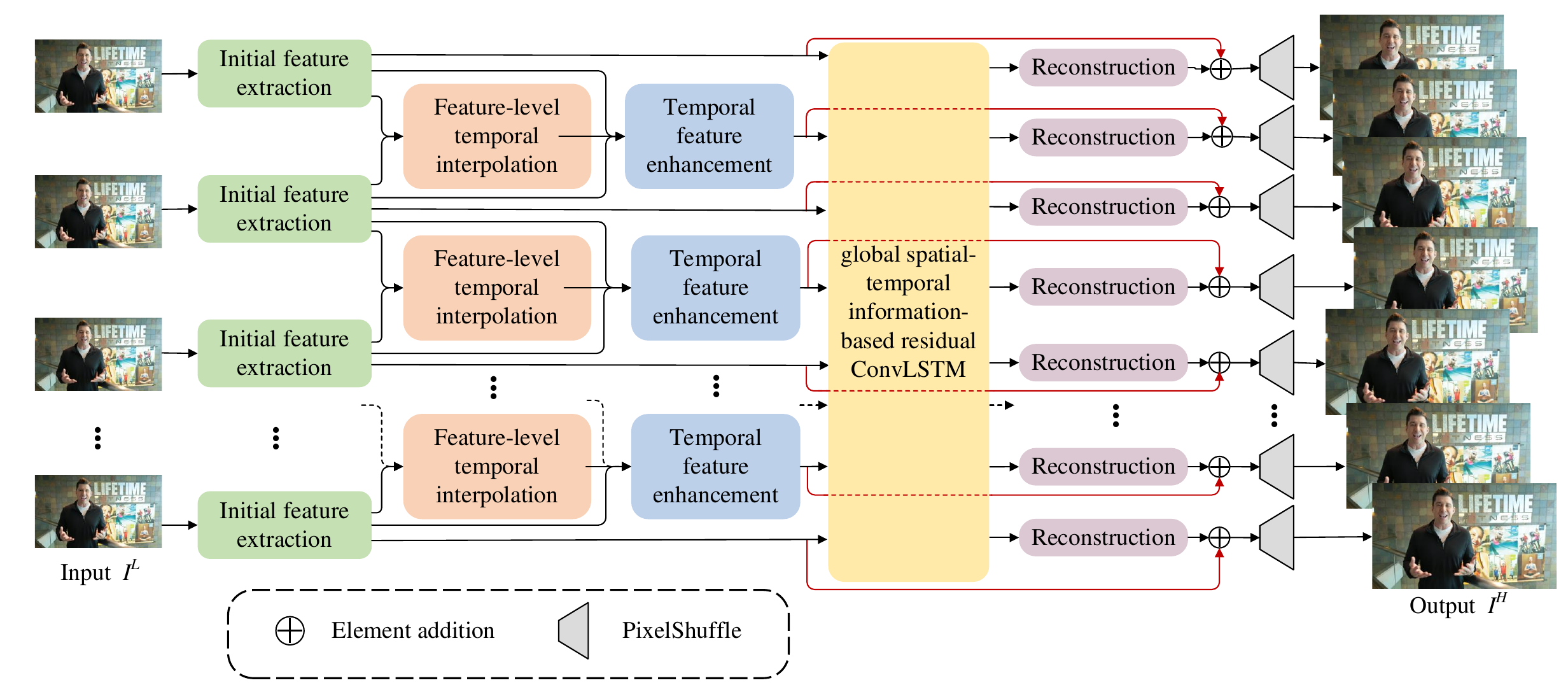}
	\caption{Overview of the proposed methods for space-time VSR.}
	\label{Overview}	
\end{figure*}

To effectively exploit the correlation between the temporal and spatial dimensions in videos, we propose an efficient convolutional long short-term memory (convLSTM)-based neural network (Fig.~\ref{Overview}). The proposed method consists of five parts: initial feature extraction, feature-level temporal interpolation, temporal feature enhancement, global spatial-temporal information-based residual convLSTM, and high-resolution reconstruction. 
Temporal interpolation uses deformable convolution to extract the forward and backward motion information at the feature level. Global spatial-temporal information is extracted and used as the initial cell state of the residual ConvLSTM.
%To generate highly accurate features, the network combines a feature-level temporal interpolation module with deformable convolutions and a residual convLSTM based on global spatial-temporal information.
Specifically, 
%in the initial feature extraction module, the initial features of the input video are extracted through the attention mechanism and multiple residual convolution blocks.
in the feature-level temporal interpolation module, deformable convolutions are used to adapt to the deformation and scale changes of objects at different scene locations. Therefore, the network effectively uses forward and backward feature information by concatenating the two features in different order to determine the inter-frame offset, directly generating more accurate interpolated frame features. 
%The temporal feature enhancement module is mainly to further enhance the accuracy of the intermediate frame features generated by the previous module.
This global spatial-temporal information-based residual convLSTM module comprises two convLSTMs. The first convLSTM computes global spatial-temporal information of the input features. The second convLSTM uses this information as its initial unit state.  Moreover, residual connections are used to retain the spatial information of each video frame.
%Finally, the high-resolution reconstruction module is to improve the spatial resolution.
The main contributions of this paper are as follows.

\begin{itemize}
	\item We propose GIRNet, a VSTSR network that combines feature-level temporal interpolation and a global spatial-temporal information-based residual convLSTM. GIRNet can better exploit spatial and temporal correlations than previous VSTSR methods. 
	\item In the feature-level temporal interpolation module, deformable convolution is strategically adopted. This choice enables the effective integration of both forward and backward feature information to derive the inter-frame offset. This offset is subsequently leveraged to directly generate interpolated frame features, facilitating a more robust representation of moving objects within the scene.
	\item In the residual convLSTM, a first convLSTM generates global spatial-temporal information from the input features. A second convLSTM uses the computed information as its initial cell state. Moreover, this second convLSTM uses residual connections to preserve the spatial features. This approach significantly enhances the accuracy of the output features.
	\item Extensive experiments demonstrate that our method outperforms the state-of-the-art.
	
\end{itemize}

The remainder of this paper is organized as follows. Related work is presented in Section II. The proposed network is described in detail in Section III. Experimental results, including an ablation study, are given in Section IV. Finally, Section V concludes the paper.

\section{Related Work}

VSR can be classified into three categories: VSSR, VTSR and VSTSR.
VSSR improves the quality of low-resolution videos by upscaling them to higher spatial resolutions. 
VTSR mitigates motion artifacts and reduces judder by interpolating frames between existing ones, thereby enhancing the overall video playback quality. 
VSTSR enhances both spatial and temporal resolutions, presents a more challenging problem.

\subsection{Video Spatial Super-Resolution}
VSSR refers to the process of transforming low-resolution videos into high-resolution videos in the spatial domain. 
Most conventional VSSR methods use four stages: feature extraction, alignment, fusion and reconstruction.
Haris \textit{et al.}~\cite{rbpn} proposed to achieve spatial super-resolution for video by incorporating convolutional neural network (CNN)-based single image super-resolution and optical flow estimation. 
Tian \textit{et al.}~\cite{tdan} proposed a temporal deformable alignment network (TDAN) to adaptively align the reference frame with each supporting frame at feature level without optical flow estimation. 
Wang \textit{et al.}~\cite{edvr} proposed a video restoration framework, namely EDVR. It consists of pyramid, cascading, and deformable (PCD) alignment modules to handle large motions, and a temporal spatial attention (TSA) fusion module to emphasize important features for subsequent restoration. 
Chan \textit{et al.}~\cite{basicvsr} proposed a succinct pipeline, namely BasicVSR, which uses bidirectional propagation to maximize information gathering. 
Wen \textit{et al.}~\cite{wen2022video} proposed a VSSR network that dynamically generates spatially adaptive filters to improve temporal alignment. 
Recently, transformers have also been used for VSSR.
Liu \textit{et al.}~\cite{liu2022learning} proposed a novel trajectory-aware
transformer for VSSR, namely TTVSR.
Qiu \textit{et al.}~\cite{qiu2023learning} proposed FTVSR++, a novel degradation-robust Frequency-Transformer for VSSR, which operates in a combined space-time-frequency domain. It distinguishes real visual texture from artifacts and incorporates dual frequency attention.

\subsection{Video Temporal Super-Resolution}
VTSR is a motion estimation-based technique that estimates the pixel values of intermediate frames by analyzing the motion between adjacent frames, thereby achieving an increase in frame rate. This method can be used to improve video smoothness. 
Common VTSR algorithms include interpolation kernel-based methods and motion compensation-based methods.
With the introduction of CNN-based optical flow algorithms~\cite{dosovitskiy2015flownet}, several algorithms using optical flow have been developed. 
Xue \textit{et al.} ~\cite{xue2019video} used the bi-directional flow to warp input frames using the backward warping function. 
Liu \textit{et al.}~\cite{liu2019deep} used cycle consistency loss to make synthesized frames more reliable as input frames. 
To deal with the occlusion problem, a common issue in optical flows, additional depth information was used to refine the optical flows in DAIN~\cite{bao2019depth}. 
Park \textit{et al.}~\cite{park2020robust} proposed a VTSR method by considering the exceptional motion patterns. 
Lee \textit{et al.}~\cite{lee2020adacof} proposed a warping module, namely Adaptive Collaboration of Flows (AdaCoF), to estimate both kernel weights and offset vectors for each target pixel to synthesize the missing frame.
Kong \textit{et al.}~\cite{kong2022progressive} introduced a progressive motion context refine network for efficient frame interpolation, which jointly predicts motion fields and image context. 
Zhu \textit{et al.}~\cite{zhu2023mfnet} proposed a frame interpolation network (namely MFNet) that focuses on motion regions. 
Liu \textit{et al.}~\cite{liu2023ttvfi} proposed a trajectory-aware transformer for VTSR. The network aims to reduce the distortion and blur resulting from inconsistent motion and inaccurate warping. 
Plack \textit{et al.}~\cite{plack2023frame} presented a transformer-based method that estimates both the interpolated frame and its expected error. 

\subsection{Video Space-Time Super-Resolution}

VSTSR aims to increase the spatial and temporal dimensions of low frame-rate and low-resolution videos.
An intuitive way is to apply VSSR and VTSR alternately. However, this approach treats, space and time independently, limiting the performance. Another way is to simultaneously generate the output video with high resolution and high frame rate. Due to missing pixels and frames in low spatial and temporal resolution videos, VSTSR is a highly ill-posed inverse problem. 
Shechtman \textit{et al.}~\cite{shechtman2005space} first proposed a VSTSR framework by using multiple LR videos of the same dynamic scene. Unlike the VTSR methods mentioned above, their method explicitly deals with the motion blur to generate sharp interpolated frames.
Shahar \textit{et al.}~\cite{shechtman2011} extended the work in~\cite{shechtman2005space} with a method that leverages the statistical recurrence of small space-time patches in a single natural video sequence.
Takeda \textit{et al.}~\cite{takeda2010spatiotemporal} considered both local spatial orientations and local motion vectors and adaptively constructed a suitable filter at every position of interest. 
Kang \textit{et al.}~\cite{kang2020deep} introduced a weighting scheme to effectively fuse all input frames without requiring explicit motion compensation. 
Dutta \textit{et al.}~\cite{dutta2021efficient} used a quadratic model to interpolate in LR space. They also reused the flowmaps and blending mask to synthesize the interpolated frame in HR space with bilinear upsampling. 
Xiang \textit{et al.}~\cite{xiang2020zooming} proposed a feature temporal interpolation network to capture local temporal contexts when interpolating LR frame features.
Haris \textit{et al.}~\cite{haris2020space} developed a deep neural network that uses direct lateral connections between multiple resolutions to present rich multi-scale features during training. 
Xu \textit{et al.}~\cite{xu2021temporal} proposed TMNet, which uses temporal modulation blocks for accurate frame interpolation and incorporates locally-temporal feature comparison modules and bi-directional deformable convLSTM for effective motion cue extraction.
Zhang \textit{et al.}~\cite{zhang2022cross} proposed a cross-frame transformers instead of traditional convolutions by dividing the input feature sequence into query, key, and value matrices to capture the spatial and temporal correlation effectively. 
Fu \textit{et al.}~\cite{10466790} proposed a generative adversarial network-based three-dimensional attention mechanism (3DAttGAN). The discriminative network uses a two-branch structure to handle the intra-frame texture details and inter-frame motion occlusions in parallel.
Xiao \textit{et al.}~\cite{XIAO2022102731} introduced a joint framework for enhancing the spatial and temporal resolution of satellite video using a feature interpolation module and a multi-scale spatial-temporal transformer.
Hu \textit{et al.}~\cite{hu2023store} developed a store-and-fetch network by effectively learning long-range spatial-temporal correlations, using backward and forward recurrent modules to store and retrieve past and future super-resolution information.
Hu \textit{et al.}~\cite{hu2023cycmunet+} proposed a cycle-projected mutual learning network (CycMuNet+) that uses iterative up- and down projections to fuse spatial and temporal features.
However, these approaches do not comprehensively exploit the innate temporal and spatial characteristics embedded within video frames.

\section{Proposed Method}

To effectively exploit the correlation between the temporal and spatial information and improve the quality of space-time super-resolution, we propose GIRNet, an end-to-end convLSTM-based neural network. Our network directly learns to map an input video with low spatial and temporal resolution to an output video with high spatial and temporal resolution.
Subsection III.A gives an overview of GIRNet. Subsection III.B introduces the feature-level temporal interpolation (FLTI) module. Subsection III.C presents the temporal feature enhancement (TFE) module. Section III.D presents the global spatial-temporal information-based residual convLSTM (GSTIR), which is the core module of our network. Finally, Subsection III.E describes the high-resolution reconstruction process with PixelShuffle.

%We overview GIRNet in subsection A. Then, we introduce novelty feature-level temporal interpolation (FLTI) in subsection B and temporal feature enhancement (TFE) in subsection C. Subsequently, we present the core module that the global spatial-temporal information-based residual convLSTM (GSTIR) in subsection D. Finally, the high-resolution reconstruction with the PixelShuffle procedure is presented in subsection E.

\subsection{Network Overview}

As illustrated in Fig.~\ref{Overview}, GIRNet consists of five parts: initial feature extraction, FLTI, TFE, GSTIR, and high-resolution reconstruction.
Specifically, we first extract the features of the input video frames through attention and multiple residual convolution blocks, respectively. 
These extracted initial features are then sent to the FLTI module to generate the features of the interpolated frames by using deformable convolution to get the inter-frame offset. 
Next, the interpolated frame features and extracted features of the input frames are fed together to the TFE module to further enhance the features of the interpolated frame. Then, the input frames and the enhanced interpolated frame features are fed into the GSTIR module for fusion and refinement. In the GSTIR module, residual connection is additionally used to retain the spatial information. The output features of the GSTIR module are then respectively regulated by residual convolution blocks in the reconstruction module. Finally, PixelShuffle operations are used to obtain space-time super-resolved video frames.
To preserve detail information, we use a global residual connection, that is, the input features of the GSTIR module are added to the output of the reconstruction module to preserve detail information, as shown by red connecting line in Fig.~\ref{Overview}.
In the following, we briefly explain the network flow with input $\mathcal{V}^L=\left\{\boldsymbol{I}_{t-1}^L, \boldsymbol{I}_{t+1}^L\right\}$ and output $\mathcal{V}^H=\left\{\boldsymbol{I}_{t-1}^H, \boldsymbol{I}_{t}^H, \boldsymbol{I}_{t+1}^H\right\}$ as an example.

\subsubsection{initial feature extraction}
\begin{figure}[htbp]
	\centering
	\includegraphics [scale=0.5]{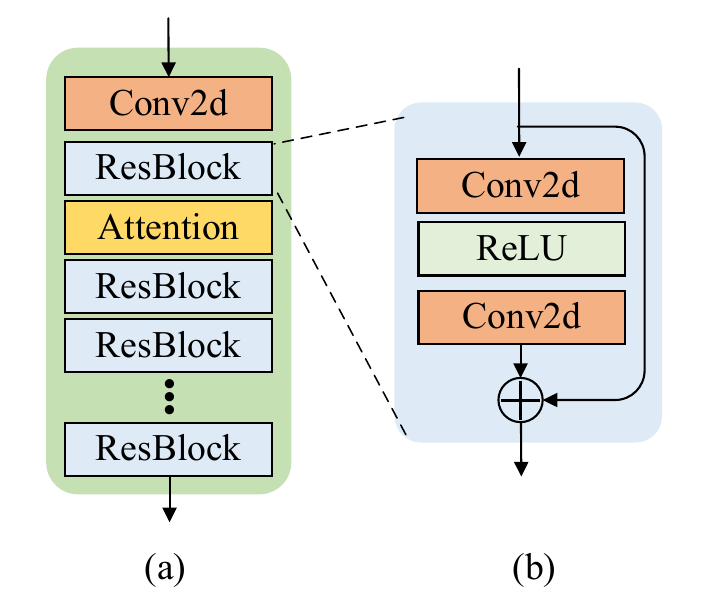}
	\caption{Initial feature extraction module and reconstruction module. }
	\label{ini}	
\end{figure}

Given the input video frames with low-frame-rate and low-resolution $\mathcal{V}^L=\left\{\boldsymbol{I}_{t-1}^L, \boldsymbol{I}_{t+1}^L\right\}$,
GIRNet first extracts the corresponding initial features $ \left\{\boldsymbol{F}_{t-1}^L, \boldsymbol{F}_{t+1}^L\right\} $, as shown in Fig.~\ref{ini} (a). 
The input video frames are passed through a 2D convolutional layer for shallow feature extraction, followed by a series of residual convolutional blocks (ResBlocks) for further feature extraction with attention to the region of interest (Fig.~\ref{ini} (b)). Here, the existing attention modules are used directly. The effect of different attention mechanisms will be studied in Section IV.C.
The number of ResBlocks is discussed in Section IV.C.1.
The initial feature extraction phase focuses on capturing shallow features from the input video frames, containing abundant pixel-level information and resulting in a smaller overlap area for receptive fields, thereby enabling the network to extract finer details.

\subsubsection{feature-level temporal interpolation}

%For video sequences, the estimation of motion information is very important. 
In this module, to get the initial intermediate frame feature $ \tilde{\boldsymbol{F}_{t}^L} $ from the extracted features $ \left\{\boldsymbol{F}_{t-1}^L, \boldsymbol{F}_{t+1}^L\right\} $,
we use deformable convolutions to exploit forward and backward information by concatenating the two features in different order, as shown in Fig.~\ref{DConv}.
During the procedure, the forward and backward information are split into two branches to calculate their offsets, respectively. Consequently, the output of the two branches are weighted together to obtain $ \tilde{\boldsymbol{F}_{t}^L} $.

\subsubsection{temporal feature enhancement}

In this module, we further enhance the initial intermediate frame feature $ \tilde{\boldsymbol{F}_{t}^L} $ to obtain a more accurate intermediate frame feature $ \boldsymbol{F}_{t}^L $. 
We again use the feature information of the forward and backward features with the initial intermediate frame feature for feature alignment and enhancement. Specifically, the enhanced intermediate frame features are obtained by residual connection.                                                                  

\subsubsection{global spatial-temporal information-based residual ConvLSTM}

LSTM can effectively convey and express features in a long term series without neglecting useful information. In this module, the features $ \left\{\boldsymbol{F}_{t-1}^L, \boldsymbol{F}_{t}^L, \boldsymbol{F}_{t+1}^L\right\} $ are used to calculate spatial-temporal global information by one convLSTM. This global information is used as the initial cell state of another convLSTM. This process generates the enhanced features $ \left\{\boldsymbol{H}_{t-1}, \boldsymbol{H}_{t}, \boldsymbol{H}_{t+1}\right\} $, which can effectively improve the accuracy of the features. 
Moreover, a residual connection from the input to the output of convLSTM is proposed to preserve spatial information.

\subsubsection{high-resolution reconstruction}
The output of the GSTIR module, i.e., $ \left\{\boldsymbol{H}_{t-1}, \boldsymbol{H}_{t}, \boldsymbol{H}_{t+1}\right\} $ is divided into three branches for high-resolution reconstruction. Both ResBlocks and an attention mechanism are used to refine the features. 
Subsequently, we increase the spatial resolution of these improved features by using PixelShuffle~\cite{pixelshuffel} and obtain the final high-frame-rate and high-resolution video sequence $\mathcal{V}^H=\left\{\boldsymbol{I}_{t-1}^H, \boldsymbol{I}_{t}^H, \boldsymbol{I}_{t+1}^H\right\}$.

\subsection{Feature-Level Temporal Interpolation}

Due to the pronounced deformations of moving objects, accurate motion estimation stands as an essential requirement in VSTSR.
Traditional convolution methods, in their fixed-receptive-field nature, offer uniform coverage of activation units. 
However, as dynamic scenes contain objects with varying scales and deformations across spatial locations, the neural network needs to adapt the scale or receptive field size for precise object localization. 
Deformable convolutions offer a powerful alternative by using deformable convolution kernels, thus facilitating adaptive feature learning.
This approach offers several advantages, including an expanded receptive field and heightened resistance to spatial transformations. By introducing a learned offset, deformable convolution empowers the convolution kernel to traverse sample points within the input feature map. This directs its focus towards regions of interest or specific targets, enhancing image quality and detail restoration in space-time super-resolution tasks. 

\begin{figure}[htbp]
	\centering
	\includegraphics [width=\linewidth]{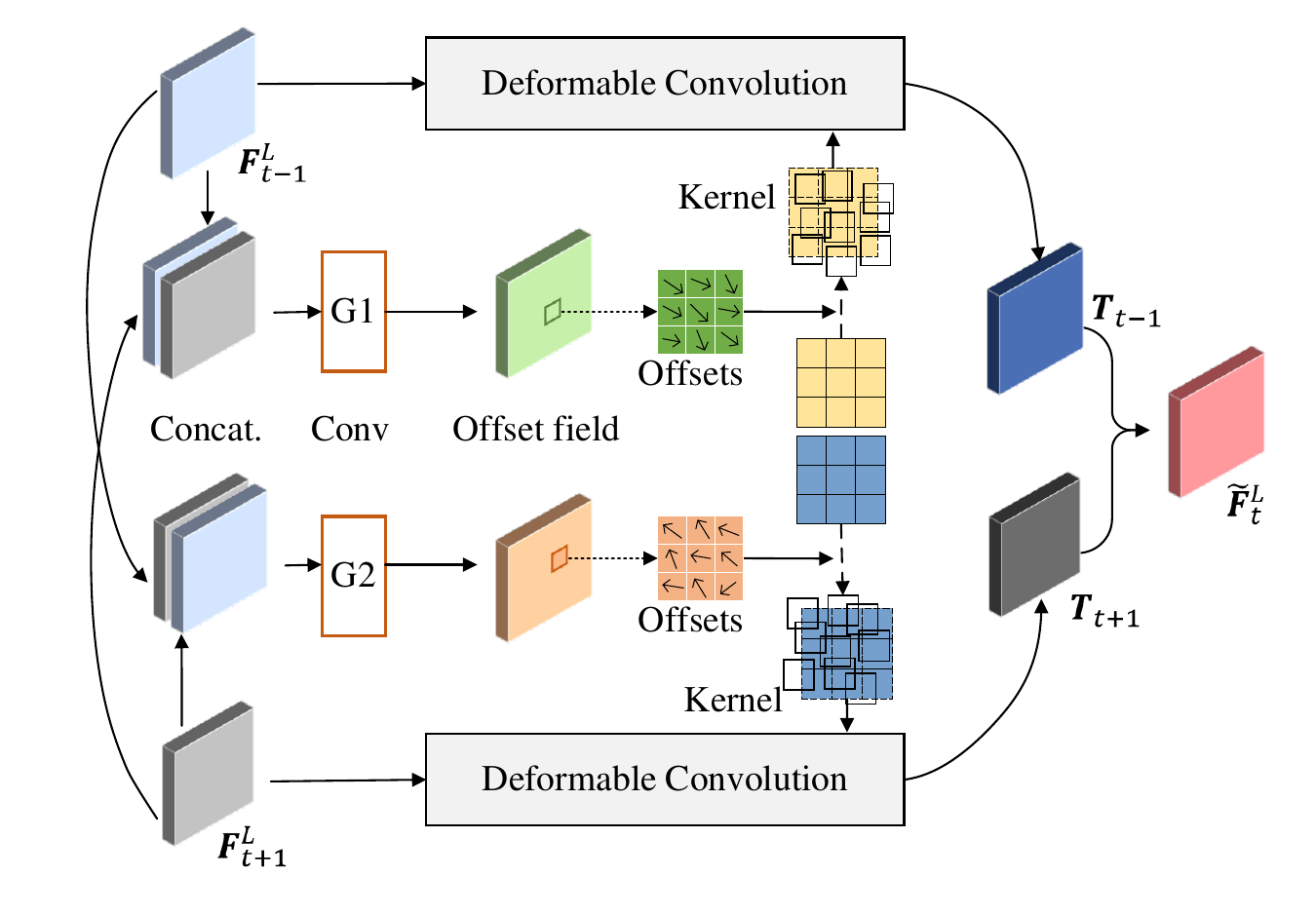}
	\caption{Feature-level temporal interpolation module. }
	\label{DConv}	
\end{figure}

Within the feature-level temporal interpolation module, our method harnesses deformable convolution effectively to exploit both forward and backward feature information for the efficient determination of the offset by concatenating the two features in different order. This strategy allows for the direct generation of interpolated frame features, as illustrated in Fig.~\ref{DConv}.

First, we concatenate the input features $ \left\{\boldsymbol{F}_{t-1}^L, \boldsymbol{F}_{t+1}^L\right\} $ according to the forward and backward arrangement.
Then, we use convolutional layers to compute two offsets 
%It takes LR frame feature maps $ \boldsymbol{F}_{1}^L $ and $ \boldsymbol{F}_{3}^L $ as input to predict an offset.
\begin{equation}
	\Delta p_1=g_1\left(\left[\boldsymbol{F}_{t-1}^L, \boldsymbol{F}_{t+1}^L\right]\right),
\end{equation}
\begin{equation}
	\Delta p_3=g_3\left(\left[\boldsymbol{F}_{t+1}^L, \boldsymbol{F}_{t-1}^L\right]\right),
\end{equation}
%where $ \Delta p_1 $ and $ \Delta p_3 $ are the learnable offsets, respectively;
where $ g_1 $ and $ g_3 $ denote convolution layers; and $ [, ] $ denotes channel-wise concatenation.
The two offsets $ \Delta p_1 $ and $ \Delta p_3 $ are implicitly learned to capture the forward and backward motion information, respectively. The offsets can enforce the synthesized LR feature map to be close to the real intermediate LR feature map.

Next, a deformable operation is applied to $\boldsymbol{F}_{t-1}^L$ and $ \boldsymbol{F}_{t+1}^L$ to obtain the features $ \boldsymbol{T}_{t-1}$ and $ \boldsymbol{T}_{t+1} $.

\begin{equation}
	T_{t-1}=\operatorname{DConv}\left(\boldsymbol{F}_{t-1}^L, \Delta p_1\right),
\end{equation}
\begin{equation}
	T_{t+1}=\operatorname{DConv}\left(\boldsymbol{F}_{t+1}^L, \Delta p_3\right).
\end{equation}

Finally, to blend the two features, we use a simple linear blending function:

\begin{equation}
	\begin{aligned}
	\tilde{\boldsymbol{F}}_t^L
	& =f\left(\boldsymbol{F}_{t-1}^L, \boldsymbol{F}_{t+1}^L\right)=H\left(T_{t-1}, T_{t+1}\right) \\
	& =\alpha * T_{t-1}+\beta * T_{t+1},
    \end{aligned}
\end{equation}
where $ H(\cdot) $ is a blending function to aggregate sampled features, $\alpha$ and $\beta$ are two learnable $ 1 \times 1 $ convolution kernels, and $ * $ denotes the convolution operator.

\subsection{Temporal Feature Enhancement}

It is essential to maintain long-term temporal consistency for each frame. For this purpose, we propose a temporal feature enhancement module to exploit the complementary information (e.g., motion cues) from adjacent frames.

\begin{figure}[htbp]
	\centering
	\includegraphics [width=\linewidth]{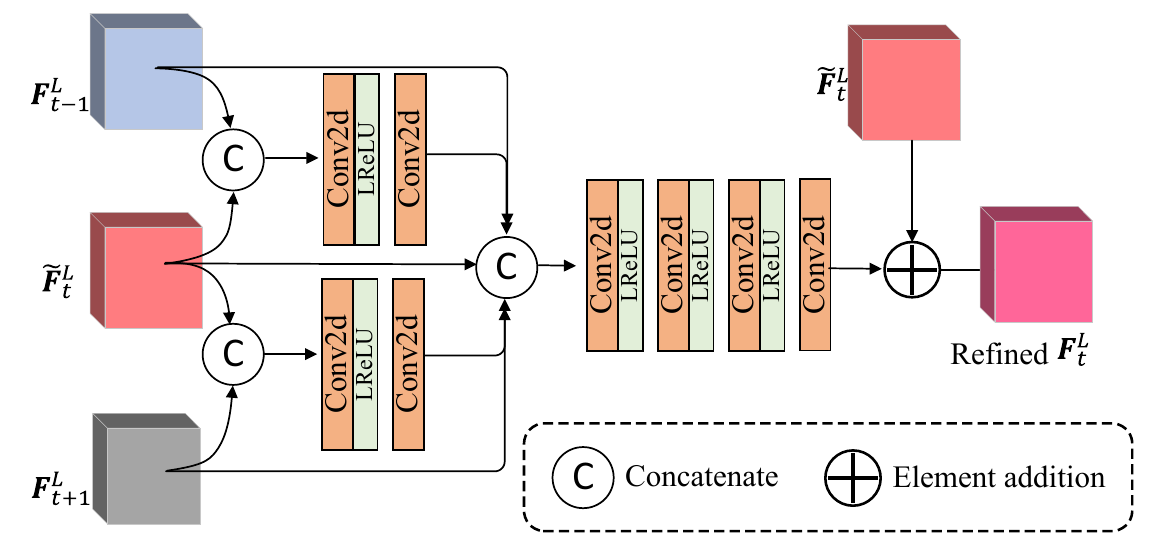}
	\caption{Temporal feature enhancement module. }
	\label{TFE}	
\end{figure}

As illustrated in Fig.~\ref{TFE}, to refine the feature map $ \tilde{\boldsymbol{F}}_t^L $ of the current frame from the adjacent feature maps $ \boldsymbol{F}_{t-1}^L $ and $ \boldsymbol{F}_{t+1}^L $, we concatenate the current frame $ \tilde{\boldsymbol{F}}_t^L $ and the adjacent frames $ (\boldsymbol{F}_{t-1}^L, \boldsymbol{F}_{t+1}^L )$ respectively and use two convolutional layers to extract the features for the forward and the backward motion information.
%Note that we learn two offsets to describe the motion cues in the forward and the backward directions. Then, the learned offset from the forward (or backward) direction is used to align the feature map $ \boldsymbol{F}_{t-1}^L $ ($ \boldsymbol{F}_{t+1}^L $ )of the previous (next) frame with that of the current frame, via one deformable convolutional layer.
Next, we concatenate the two adjacent frames feature, the current frame feature and the forward and backward motion features. Following this, we conduct feature comparison using four $1 \times 1$ convolutional layers and an addition operation. As a result, we get a refined feature $ \boldsymbol{F}_t^L $. The extra motion information provides more accurate guidance for the generation of intermediate interpolation features.

\subsection{Global Spatial-Temporal Information-based Residual ConvLSTM}
%Now we have consecutive frame features to generate the corresponding HR video frames. 
As temporal information is also vital in spatial reconstruction~\cite{tao2017detail}, we aggregate temporal contexts of neighboring frames using convLSTM~\cite{shi2015convolutional}.

\begin{figure}[htbp]
	\centering
	\includegraphics [width=\linewidth]{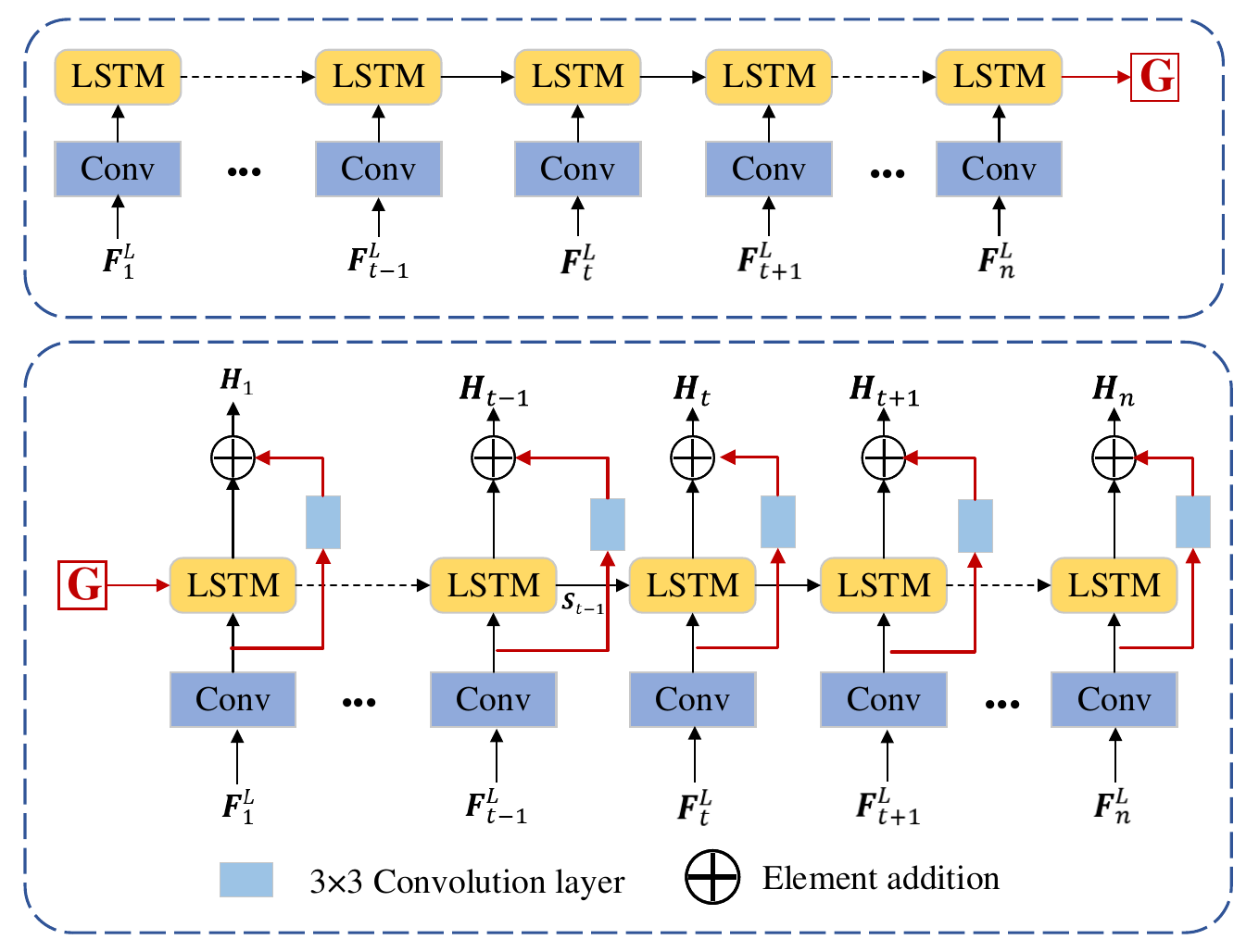}
	\caption{Global spatial-temporal information-based residual convLSTM: the top section generates global spatial-temporal information ($G$), while the bottom section generates output features through the utilization of residual connections.}
	\label{LSTM}	
\end{figure}

In convLSTM, the initial cell state plays an important role. We first calculate the global spatial-temporal information as the initial cell state $G$, based on the features of consecutive frames, as shown in the top part of Fig.~\ref{LSTM}. This is the core part of this module. The global spatial-temporal information can provide useful guidance for the accuracy of features in each frame.
This convLSTM is many-to-one, that is, there are $ n $ inputs, $ \left\{\boldsymbol{F}_{1}^L, \dots ,\boldsymbol{F}_{t-1}^L, \boldsymbol{F}_{t}^L, \boldsymbol{F}_{t+1}^L,\dots,\boldsymbol{F}_{n}^L\right\}$, and only one output, $ G $. $ G $ synthesizes the input information of $ n $ frames as processed by the neural network:
\begin{equation}
	G = ConvLSTM(\boldsymbol{F}_{1}^L, \dots ,\boldsymbol{F}_{t-1}^L, \boldsymbol{F}_{t}^L, \boldsymbol{F}_{t+1}^L,\dots,\boldsymbol{F}_{n}^L).
\end{equation}

To make the obtained features more accurate, we design another many-to-many convLSTM. The input are the features $ \left\{\boldsymbol{F}_{1}^L, \dots ,\boldsymbol{F}_{t-1}^L, \boldsymbol{F}_{t}^L, \boldsymbol{F}_{t+1}^L,\dots,\boldsymbol{F}_{n}^L\right\}$ obtained from the previous module, and the output are the corresponding enhanced features $ \left\{\boldsymbol{H}_{1}, \dots ,\boldsymbol{H}_{t-1}, \boldsymbol{H}_{t}, \boldsymbol{H}_{t+1},\dots,\boldsymbol{H}_{n}\right\}$. 
Here, we use the previously computed $ G $ as the initial cell state of convLSTM to guide the output generation.
Moreover, we add a $ 3 \times 3 $ convolution layer as residual connection between the input and output of convLSTM to preserve the spatial feature of each frame (see the bottom part of Fig.~\ref{LSTM}),

\begin{equation}
	\begin{aligned}
	\boldsymbol{H}_t=&ConvLSTM\left(\operatorname{Conv}\left(\boldsymbol{F}_t^L\right), \boldsymbol{S}_{t-1}\right)
	\\&+\operatorname{Conv}\left(\operatorname{Conv}\left(\boldsymbol{F}_t^L\right)\right),
	\end{aligned}
\end{equation}

\begin{equation}
	\boldsymbol{S}_{t-1}=ConvLSTM\left(\operatorname{Conv}\left(\boldsymbol{F}_{t-1}^L\right)\right).
\end{equation}

\subsection{High-Resolution Reconstruction}

\begin{figure}[htbp]
	\centering
	\includegraphics [scale=0.5]{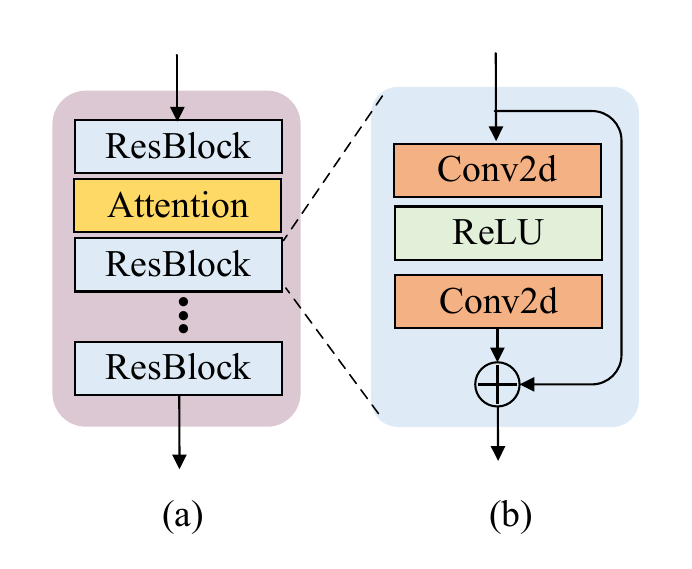}
	\caption{Initial feature extraction module and reconstruction module. }
	\label{Rec}	
\end{figure}

We proceed with spatial refinement for the features using ResBlocks with attention to get the refined features $ \left\{\boldsymbol{F}_{i}^H, i=1,2,...,n\right\} $, as shown in Fig.~\ref{Rec}. 
%The number of ResBlocks is discussed in Section IV.C.1. 
Finally, we feed $ \boldsymbol{F}_{i}^H $ into the PixelShuffle layers to reconstruct HR video frames $\mathcal{V}^H=\left\{\boldsymbol{I}_{i}^H, i=1,2,...,n\right\}$.
The PixelShuffle layer can effectively preserve the detail of the image when increasing the resolution. This helps to generate sharper and clearer images without introducing too much blur.

To optimize our network, we use the reconstruction loss function
\begin{equation}
	%l_{rec}=\sqrt{\left\|I_t^{GT}-I_t^H\right\|^2+\epsilon^2},
	l_{rec}= \sqrt{\frac{1}{N} {\textstyle \sum_{t=1}^{N}}\left (I_{t}^{GT}-I_{t}^{H}\right ) ^{2}+\epsilon ^{2}} ,
\end{equation}
where $ I_t^{GT} $denotes the $ t $-th ground-truth HR video frame, $N$ is the total number of video frames, and $ \epsilon $ is empirically set to $ 10^{-3} $ based on the Charbonnier penalty function~\cite{lai2017deep}.

\section{Experimental Results and Analysis}

\subsection{Datasets and Experimental Settings}

To assess the performance of the proposed method, we used the Vimeo90K dataset~\cite{xue2019video}. The training set consists of 64,612 videos, and the test set consists of 7,824 videos, each with dimension 7 (frames)$ \times $ 448 (spatial width) $ \times $ 256 (spatial height). 
To measure the performance under different motion conditions, we split the Vimeo90K test set into three groups: fast motion, medium motion, and slow motion, as in~\cite{xue2019video}. 

We generated low-resolution videos by spatially and temporally sub-sampling. Specifically, the spatial resolution was down-sampled through bi-cubic interpolation, while the even frames were removed to reduce the temporal resolution. The peak signal-to-noise ratio (PSNR) and structural similarity index (SSIM)~\cite{wang2004image} were adopted to evaluate the quality of the generated full-resolution videos.
%to 112$ \times $64 

The proposed network was implemented on the PyTorch platform and trained using a graphical card with NVIDIA 4090 GPU. In the implementation, we randomly cropped the videos for training to patches of size 32 $ \times $ 32$ \times $ 4 as input and used the corresponding ground truth video patch as the labels. During training, we set the batch size to 8 and used the Adam optimizer with $ \beta _1=0.5 $ and $ \beta _2=0.99 $. We initialized the learning rate at 0.0001 and decreased it by a factor of 0.5 every 60 epochs.

\subsection{Comparison with State-of-the-Art Methods}

We compared the proposed GIRNet with state-of-the-art two-stage and one-stage VSTSR methods. 
For two-stage VSTSR methods, we combined state-of-the-art VSSR and VTSR methods to obtain the VSTSR videos as anchors. Specifically, DBPN~\cite{haris2018deep}, RBPN~\cite{rbpn}, TDAN~\cite{tdan}, and BasicVSR~\cite{basicvsr} were used for VSSR, while ToFlow~\cite{xue2019video} and DAIN~\cite{bao2019depth} were adopted for VTSR. 
For one-stage VSTSR methods, Zooming SlowMo~\cite{xiang2020zooming}, STARnet~\cite{haris2020space}, TMNet~\cite{xu2021temporal} and 3DAttGAN~\cite{10466790} were compared.
For fairness, the comparison methods were retrained with the same Vimeo90K training set and tested with the same test set.
Note that the super-resolution factors ($\times2, \times4, \times8$) refer to spatial super-resolution, while the temporal super-resolution refers to doubling the frame rate.

\begin{figure*}[htbp]
	\centering
	\includegraphics [width=\linewidth]{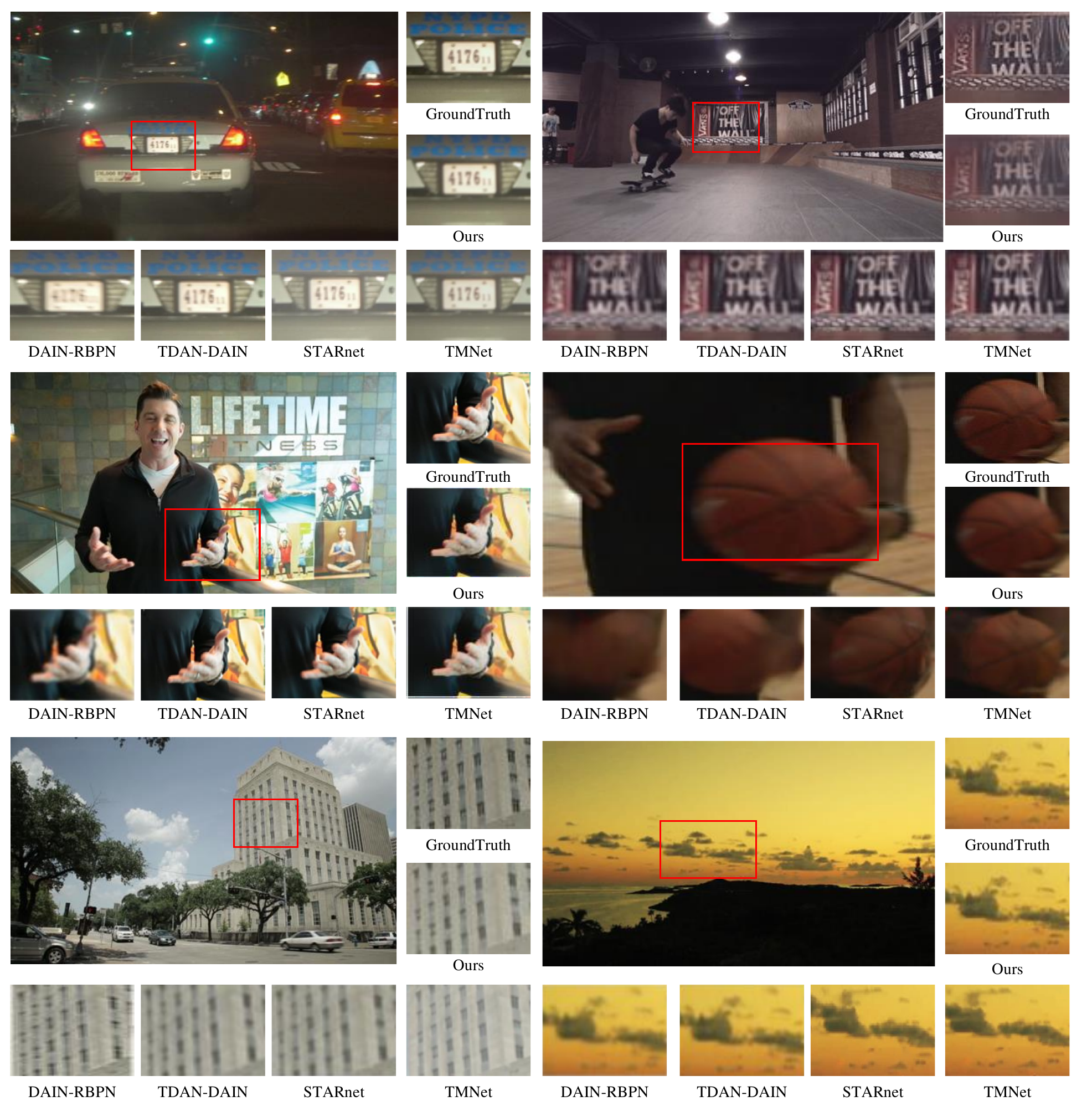}
	\caption{Visual comparison of VSTSR methods. }
	\label{Visual}
\end{figure*}

\subsubsection{Quantitative Results}

Quantitative results are presented in Table~\ref{VSTSR}. We can see that one-stage methods significantly outperformed the two-stage methods in all metrics. 
Specifically, the combination VSSR+VTSR performed better than VTSR+VSSR.
The performance of the best two-stage method (BasicVSR+DAIN) was much lower than that of GIRNet for $\times$4 VSTSR on the Vimeo90K dataset. 
Furthermore, GIRNet outperformed STARnet~\cite{haris2020space}, Zooming SlowMo~\cite{xiang2020zooming}, TMNet~\cite{xu2021temporal} and 3DAttGAN~\cite{10466790}. 
From Table~\ref{VSTSR}, we can see that, for $\times$4 VSTSR, GIRNet gave the highest PSNR (32.06 dB on average) and SSIM (0.951) on Vimeo90K.
We also compared the one-stage VSTSR methods with the proposed GIRNet across three super-resolution factors ($\times2, \times4, \times8$), as shown in Table~\ref{diff-SR}. We can see that GIRNet was also the best in terms of both metrics.
In addition, we tested the effectiveness of our method on REDS~\cite{son2021ntire} and UCF101~\cite{soomro2012ucf101} as shown in Table~\ref{diff-dataset}.
All these results validate the effectiveness of GIRNet for the VSTSR task.

\begin{table*}[htbp]
	\centering
	\caption{Quantitative comparisons($\times 4$) of the state-of-the-art methods for VSTSR on Vimeo90K.}
	\renewcommand\arraystretch{1.5}  
	%\resizebox{\linewidth}{!}{
			\begin{tabular}{l|c c||l|c c}
					\hline
					\makecell{Method \\ (VSSR+VTSR/VSTSR)} & PSNR↑ & SSIM↑ & 
					\makecell{Method \\(VTSR+VSSR/VSTSR)} & PSNR↑ & SSIM↑  \\
					\hline
					DBPN+ToFlow & 29.87 & 0.905 & ToFlow+DBPN & 28.82 & 0.887 \\
					RBPN+ToFlow & 30.29 & 0.913 & ToFlow+RBPN & 29.25 & 0.901 \\
					TDAN+ToFlow & 30.41 & 0.918 & ToFlow+TDAN & 29.65 & 0.908 \\
					BasicVSR+ToFlow & 30.58 & 0.924 & ToFlow+BasicVSR & 30.22 & 0.915 \\
					\hline
					DBPN+DAIN & 30.02 & 0.908 & DAIN+DBPN & 29.22 & 0.901 \\
					RBPN+DAIN & 30.25 & 0.916 & DAIN+RBPN & 29.42 & 0.906 \\
					TDAN+DAIN & 30.52 & 0.921 & DAIN+TDAN & 29.83 & 0.911 \\
					BasicVSR+DAIN & 30.72 & 0.925 & DAIN+BasicVSR &30.45& 0.919  \\
					\hline
					STARnet & 30.61  &  0.924 & Zooming  &  30.89  &  0.925 \\
					TMNet  & 30.92  &  0.928 & 3DAttGAN & 31.86  &  0.945\\
					GIRNet & \textbf{32.06}  &  \textbf{0.951} &- &- &- \\
					\hline
					%~\cite{haris2020space}~\cite{xiang2020zooming}~\cite{xu2021temporal}
				\end{tabular}%
	%	}
	\label{VSTSR}%
\end{table*}

\begin{table}[htbp]
	\centering
	\caption{Quantitative comparisons($\times 2$, $\times 4$, $\times 8$) of one-stage methods for VSTSR on Vimeo90K.}
	\renewcommand\arraystretch{1.5}  
	\resizebox{\linewidth}{!}{
		\begin{tabular}{l|c c|c c|c c}
			\hline
			Methods & PSNR↑ & SSIM↑ & PSNR↑ & SSIM↑ & PSNR↑ & SSIM↑ \\
			VSTSR & $\times 2$ & $\times 2$ & $\times 4$ & $\times 4$ & $\times 8$ & $\times 8$\\
			\hline
			STARnet &33.04& 0.935&	30.61&	0.924&	26.36&	0.834 \\
			Zooming &33.27&	0.963&	30.89&	0.925&	26.83&	0.851\\
			TMNet  & 33.30&	0.964&	30.92&	0.928&	27.00&	0.854\\
			GIRNet  &\textbf{34.02}&	\textbf{0.971}&\textbf{32.06}&	\textbf{0.951}&	\textbf{27.96}&	\textbf{0.882}\\
			\hline
		\end{tabular}%
	}
	\label{diff-SR}%
\end{table}

\begin{table}[htbp]
	\centering
	\caption{Quantitative comparisons ($\times4$) of one-stage methods for VSTSR on REDS and UCF101 datasets.}
	\renewcommand\arraystretch{1.5}  
%	\resizebox{\linewidth}{!}{
		\begin{tabular}{l|c c|c c}
			\hline
			\multirow{2}{*}{Methods} & \multicolumn{2}{c|}{ REDS }& \multicolumn{2}{c}{ UCF101 }\\
			\cline{2-5}
			 & PSNR & SSIM & PSNR & SSIM  \\
			\hline
			STARnet &28.48 & 0.881 & 28.83&	0.920 \\
			Zooming &28.71 &0.884&  28.93&	0.923   \\
			TMNet &28.73 & 0.887&   28.99	&0.924   \\
			3DAttGAN &28.81 & 0.890&   29.26	&0.928   \\
			GIRNet  &\textbf{28.95} &\textbf{0.891} & \textbf{29.63}&	\textbf{0.932}  \\
			\hline
		\end{tabular}%
%	}
	\label{diff-dataset}%
\end{table}

GIRNet can also achieve VTSR only by replacing the PixelShuffle layer with a convolutional reconstruction layer. To illustrate the VTSR performance of GIRNet, we compared it with three VTSR methods (ToFlow~\cite{xue2019video}, MEMC-Net~\cite{bao2019memc}, and DAIN~\cite{bao2019depth}), see Table~\ref{VTSR}. We can see that GIRNet outperformed them, showing that it benefited from the repeated use of the forward and backward information for alignment and the long-term temporal information.

\begin{table}[htbp]
	\centering
	\caption{Quantitative comparisons for VTSR on Vimeo90K.}
	\renewcommand\arraystretch{1.5}  
	\begin{tabular}{c|c c c c}
		\hline
		Method &   ToFlow & MEMC-Net & DAIN & GIRNet-temporal  \\
		\hline
		PSNR      &  33.73 & 34.29 & 34.71 & \textbf{35.02}  \\
		SSIM      &  0.968 & 0.974 & 0.976 & \textbf{0.979}\\
		\hline
	\end{tabular}
	\label{VTSR}
\end{table}

\begin{table}[htbp]
	\centering
	\caption{Computational complexity comparison on Vimeo90K.}
	\renewcommand\arraystretch{2}  
	\resizebox{\linewidth}{!}{
		\begin{tabular}{c|c|c|c|c|c|c}
			\hline
			Methods & \makecell{RBPN-\\DAIN} & \makecell{TDAN-\\DAIN}& STARnet &TMNet& 3DAttGAN & GIRNet \\
			\hline
			\makecell{Parameters-\\(million)}   & 36.7 & 26.2 & 111.61 &12.26 & 20.3 &12.42\\
			\hline
			FLOPs(G) & 1776 & - & 1893 & 2769 & 975 & 2064\\
			\hline
			Speed(fps)      & 4.26 & 3.52 &13.08 &14.69 & \textbf{16.98} & 15.13\\
			\hline
		\end{tabular}
	}
	\label{time}
\end{table}

\subsubsection{Qualitative Results}

The qualitative results of five VSTSR baselines are shown in Fig.~\ref{Visual}. 
The two-stage VSTSR methods tended to produce blurry results as they ignore the mutual relations between VSSR and VTSR, which help the accurate texture inference. Compared to two-stage methods, one-stage VSTSR methods generated more accurate results. 
From Fig.~\ref{Visual}, we can see that the proposed method achieved the best visual quality, especially in texture details.

\subsubsection{Computation Complexity}

%Moreover, we also investigated the model sizes and runtime of different methods in Table~\ref{time}. 
Table~\ref{time} compares the model size and runtime of various methods. 
For synthesizing high-quality frames, VSSR and VTSR networks usually need very large frame reconstruction modules. Thus, the two-stage VSTSR methods contain a huge number of parameters. At the same time, one-stage networks need fewer parameters than the two-stage networks. From Table~\ref{time}, we can see that the number of parameters of GIRNet was the second highest. For runtime, one-stage methods also run much faster than the two-stage ones. The runtime of the proposed method achieved 15.13 fps.

\subsection{Ablation Study}

To further demonstrate the effectiveness of the modules, we conducted an ablation study. Specifically, we assessed 
1) the impact of the number of ResBlocks on two key steps: initial feature extraction and high-resolution reconstruction;
2) the attention mechanism in the initial feature extraction and the influence of overall residual connections on the outcome;
3) the deformable convolution compared to an ordinary convolution;
4) the impact of long-term information as initial hidden state in the GSTIR module and the design of the residual structure in Fig.~\ref{LSTM}.
In this part, all the experimental results were based on the Vimeo90K test set.

\begin{figure*}[htbp]
	\centering
	\includegraphics [scale=0.6]{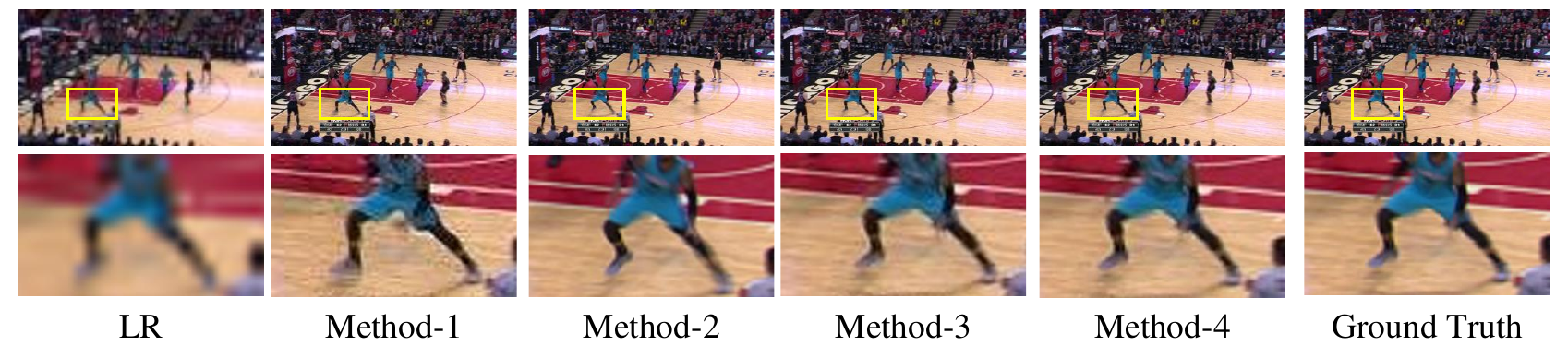}
	\caption{Visual comparisons ($\times4$) of four variants for VSTSR on Vimeo90K.}
	\label{four_variants}
\end{figure*}

\subsubsection{Effect of the Number of ResBlocks}

In the proposed method, we use ResBlocks in the initial feature extraction and in the high-resolution reconstruction.
Tables~\ref{numResBlocks-I} and~\ref{numResBlocks-H} show the impact of the number of ResBlocks on performance. 

In general, the quality can be improved by using more ResBlocks. 
For the test set of Vimeo90K, when the number of ResBlocks was increased from 3 to 9 in initial feature extraction, the PSNR improvements were 2.61 dB for Vimeo-fast, 3.26 dB for Vimeo-medium, and 2.76 dB for Vimeo-slow. 
When the number of ResBlocks was raised from 3 to 7 in high-resolution reconstruction, the PSNR improvements were 1.61 dB for Vimeo-fast, 1.47 dB for Vimeo-medium, and 1.22 dB for Vimeo-slow. 
Therefore, we opted for nine ResBlocks in the initial feature extraction and seven in the reconstruction.
%To balance the computational complexity and performance, we set the number of ResBlocks to 7 for Vid4 and 9 for Vimeo90K.

\begin{table}[htbp]
	\centering
	\caption{Impact of the number of ResBlocks in the initial feature extraction module.}
	\renewcommand\arraystretch{1.5}
	%\resizebox{\linewidth}{!}{ 
		\begin{tabular}{c|c|c|c|c|c|c}
			\hline
			\multirow{2}{*}{ResBlocks} & \multicolumn{2}{c|}{ Vimeo-fast }  & \multicolumn{2}{c|}{ Vimeo-medium }& \multicolumn{2}{c}{ Vimeo-slow }\\
			\cline{2-7}
			& PSNR & SSIM & PSNR & SSIM & PSNR & SSIM \\
			\hline
			3    & 30.45 & 0.942 & 29.37 & 0.910 & 27.79 & 0.924\\
			5    & 31.14 & 0.951 & 30.56 & 0.926 & 28.54 & 0.931\\
			7    & 32.78 & 0.959 & 31.23 & 0.953 & 29.98 & 0.945\\
			9    & \textbf{33.06} & \textbf{0.961} & \textbf{32.63} & \textbf{0.958} & \textbf{30.55} & \textbf{0.938}\\
			\hline
		\end{tabular}
		%}
	\label{numResBlocks-I}
\end{table}

\begin{table}[htbp]
	\centering
	\caption{Impact of the number of ResBlocks in the high-resolution reconstruction.}
	\renewcommand\arraystretch{1.5}  
	%\resizebox{\linewidth}{!}{
		\begin{tabular}{c|c|c|c|c|c|c}
			\hline
			\multirow{2}{*}{ResBlocks} & \multicolumn{2}{c|}{ Vimeo-fast }  & \multicolumn{2}{c|}{ Vimeo-medium }& \multicolumn{2}{c}{ Vimeo-slow }\\
			\cline{2-7}
			& PSNR & SSIM & PSNR & SSIM & PSNR & SSIM \\
			\hline
			3   &  31.45 & 0.952 & 31.16 & 0.949  & 29.33 & 0.923\\
			5   &  32.69 & 0.959 & 31.98 & 0.955  & 29.95 & 0.931\\
			7   &  \textbf{33.06} & \textbf{0.961} & \textbf{32.63} & \textbf{0.958} & \textbf{30.55} & \textbf{0.938}\\
			\hline
		\end{tabular}
		%	}
	\label{numResBlocks-H}
\end{table}

\begin{table*}[htbp]
	\centering
	\caption{Effectiveness of attention module and residual connection on Vimeo90K.}
	\renewcommand\arraystretch{1.5} 
%	\resizebox{\textwidth}{!}{ 
		\begin{tabular}{c|c|c|c|c|c|c|c|c|c}
			\hline
			\multirow{2}{*}{Methods} & \multirow{2}{*}{attention-1} & \multirow{2}{*}{attention-2} & \multirow{2}{*}{global residual connection} & \multicolumn{2}{c|}{Vimeo-fast} & \multicolumn{2}{c|}{Vimeo-medium} & \multicolumn{2}{c}{Vimeo-slow} \\
			\cline{5-10}          &       &       &       & PSNR & SSIM  & PSNR  & SSIM  & PSNR  & SSIM \\
			\hline
			Method-1 &       &       &       & 32.78 & 0.951 & 31.96  & 0.939 & 30.03 & 0.921 \\
			\hline
			Method-2 & \checkmark     &       &       & 32.96 & 0.956 & 32.37  & 0.942 & 30.25 & 0.931 \\
			\hline
			Method-3 & \checkmark     &       & \checkmark     & 33.01    & 0.959 & 32.58  & 0.946 & 30.36 & 0.934 \\
			\hline
			Method-4 &       & \checkmark     & \checkmark     & \textbf{33.06} & \textbf{0.961} & \textbf{32.63} & \textbf{0.958} & \textbf{30.55} & \textbf{0.938}\\
			\hline
		\end{tabular}%
%	}
	\label{attention}%
\end{table*}%
%%%%%%%%%%%%%%%%%%%%%%%%%%%%%%%%%%

\subsubsection{Effectiveness of Attention Module and residual Connection}

In GIRNet, we used an attention mechanism in the initial feature extraction and high-resolution reconstruction to further improve the results. 
Table~\ref{attention} shows the experimental results of the attention module. In the table, "attention 1" denotes the attention mechanism of CBAM~\cite{woo2018cbam}, while ``attention2" denotes the attention method used in Fcanet~\cite{qin2021fcanet}.
From the table, we can see that different attention mechanisms produced different results, with Fcanet~\cite{qin2021fcanet} proving more effective than CBAM.
Moreover, the inclusion of global residual connections, that is, adding the reconstruction features and the GSTIR module input features, aimed at preserving detailed information, is also reflected in the results presented in Table~\ref{attention}. Incorporating global residual connections improved the VSTSR performance.

\begin{figure}[htbp]
	\centering
	\includegraphics [scale=0.35]{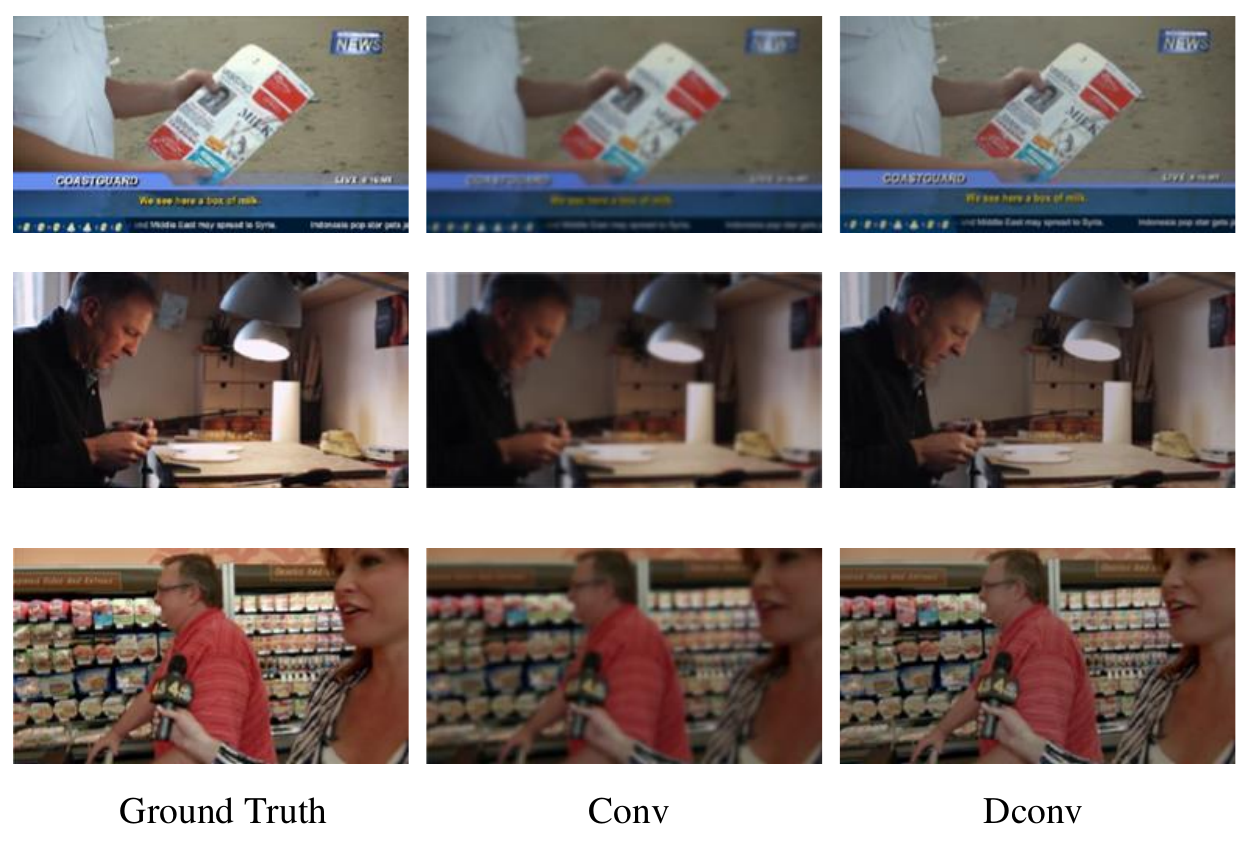}
	\caption{Visual results with and without deformable convolution.}
	\label{w/o Dconv}
\end{figure}

\begin{figure*}[htbp]
	\centering
	\includegraphics [scale=0.7]{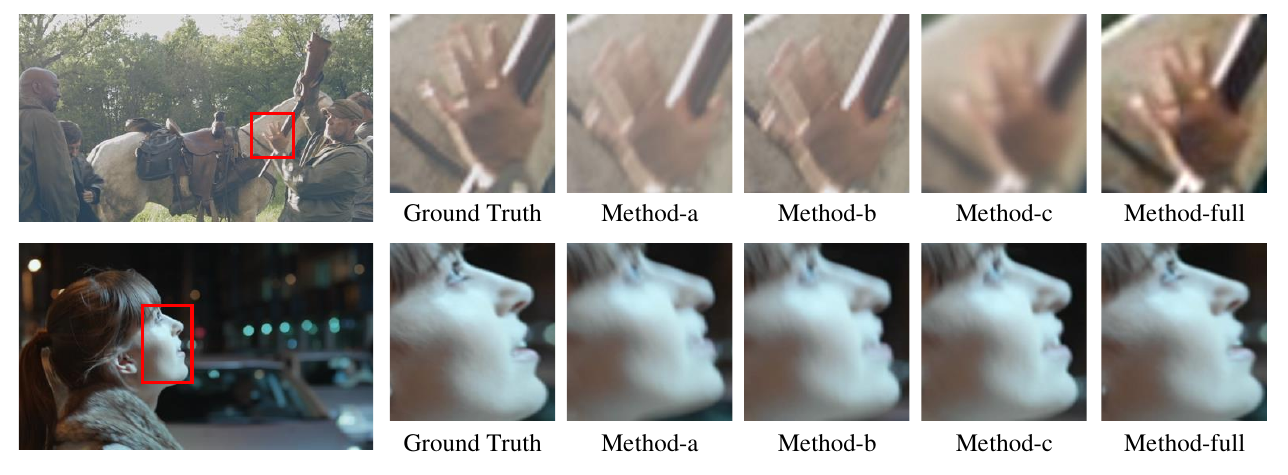}
	\caption{Visual results with global spatial-temporal information-based residual convLSTM module.}
	\label{vis-LSTM}
\end{figure*}

Fig.~\ref{four_variants} shows a visual quality comparison for attention and residual connection. We can see that the results obtained using ``attention2" were significantly better than those with ``attention1", and the global residual connection also helped to improve the subjective quality significantly.
% Table generated by Excel2LaTeX from sheet 'Sheet7'

\subsubsection{Effectiveness of Deformable Convolution}

Deformable convolution in feature-level temporal interpolation can focus on the region or object of interest and make use of the forward and backward information efficiently. 
To study its performance, we compared the results using ordinary convolution (Conv) and deformable convolution (Dconv). 
From Table~\ref{deformable_convolution}, we can see that the average PSNR increased by 0.89 dB when using deformable convolution. Fig.~\ref{w/o Dconv} shows a visual quality comparison between models using common convolution and those using deformable convolution. We can see that deformable convolution can significantly enhance subjective quality. 

\begin{table}[htbp]
	\centering
	\caption{Effectiveness of deformable convolution on Vimeo90K.}
	\renewcommand\arraystretch{1.5}  
	\resizebox{\linewidth}{!}{ 
		\begin{tabular}{c|c|c|c|c|c|c}
			\hline
			\multirow{2}{*}{Method} & \multicolumn{2}{c|}{ Vimeo-fast }  & \multicolumn{2}{c|}{ Vimeo-medium }& \multicolumn{2}{c}{ Vimeo-slow }\\
			\cline{2-7}
			& PSNR & SSIM & PSNR & SSIM & PSNR & SSIM \\
			\hline
			Conv    & 32.67 & 0.956 & 31.24 & 0.954  & 29.67 & 0.931 \\
			Dconv   &  \textbf{33.06} & \textbf{0.961} & \textbf{32.63} & \textbf{0.958} & \textbf{30.55} & \textbf{0.938}\\
			\hline
		\end{tabular}
	}
	\label{deformable_convolution}
\end{table}

\subsubsection{Effectiveness of Global Spatial-Temporal Information-based Residual ConvLSTM}

\begin{table}[htbp]
	\centering
	\caption{Effectiveness of global spatial-temporal information-based residual convLSTM module on Vimeo90K.}
	\renewcommand\arraystretch{1.5}  
	\begin{tabular}{c|c|c|c|c}
		\hline
		Methods &  \makecell{global\\ information}  & \makecell{residual\\connection} & PSNR & SSIM \\
		\hline
		Method-a &       &       & 31.17 & 0.927 \\
		\hline
		Method-b &      & \checkmark     & 31.26 & 0.931 \\
		\hline
		Method-c &  \checkmark   &  & 32.01   & 0.948 \\
		\hline
		Method-full  & \checkmark     & \checkmark     & \textbf{32.06}  &  \textbf{0.951} \\
		\hline
	\end{tabular}%
	\label{w/o convLSTM}%
\end{table}%

The GSTIR module consists of two components: generation of global spatial-temporal information and residual connections (see Fig.~\ref{LSTM}). Table~\ref{w/o convLSTM} compares four variants: removal of GSTIR from the network (Method-a), use of GSTIR without global information (Method-b), use of GSTIR without residual connections (Method-c), use of GSTIR with both global information and residual connections (Method-full). Including GSTIR in the network had a significant impact, increasing the PSNR by 0.89 dB and the SSIM by 0.024. Moreover, adding the global information was more important than adding the residual connections. 
Fig.~\ref{vis-LSTM} compares the visual quality for two videos. 
Notably, the addition of global information corrected errors in synthesis (see, e.g., how the position of the hand in the first image was corrected).

\section{Conclusion}
We presented GIRNet, a highly efficient network for VSTSR based on convLSTM. GIRNet consists of five components: initial feature extraction, feature-level temporal interpolation, temporal feature enhancement, global spatial-temporal information-based residual convLSTM, and high-resolution reconstruction. 
The successive feature-level temporal interpolation module leverages deformable convolution to compute offsets based on the input frame features. This approach significantly enhances the accuracy of the interpolated frame features. 
In the global spatial-temporal information-based residual convLSTM module, a first convLSTM generates global spatial-temporal information from the input features provided by the temporal feature enhancement module. This computed information is then used to initialize a subsequent convLSTM. We use residual connections in the second convLSTM to preserve spatial information.
Experimental results show that GIRNet outperforms state-of-the-art methods in terms of both subjective and objective quality for various super-resolution factors. 
In our future work, we aim to further enhance space-time super-resolution performance by exploring the unique characteristics of large motion scenarios. In addition, since our proposed method still cannot achieve real-time performance, further research is required to reduce its time complexity.

\bibliographystyle{IEEEtran} 
\bibliography{reference}

\begin{thebibliography}{10}
\providecommand{\url}[1]{#1}
\csname url@samestyle\endcsname
\providecommand{\newblock}{\relax}
\providecommand{\bibinfo}[2]{#2}
\providecommand{\BIBentrySTDinterwordspacing}{\spaceskip=0pt\relax}
\providecommand{\BIBentryALTinterwordstretchfactor}{4}
\providecommand{\BIBentryALTinterwordspacing}{\spaceskip=\fontdimen2\font plus
\BIBentryALTinterwordstretchfactor\fontdimen3\font minus
  \fontdimen4\font\relax}
\providecommand{\BIBforeignlanguage}[2]{{%
\expandafter\ifx\csname l@#1\endcsname\relax
\typeout{** WARNING: IEEEtran.bst: No hyphenation pattern has been}%
\typeout{** loaded for the language `#1'. Using the pattern for}%
\typeout{** the default language instead.}%
\else
\language=\csname l@#1\endcsname
\fi
#2}}
\providecommand{\BIBdecl}{\relax}
\BIBdecl

\bibitem{10420512}
Y.~Huang, J.~Huang, J.~Liu, M.~Yan, Y.~Dong, J.~Lv, C.~Chen, and S.~Chen,
  ``Wavedm: Wavelet-based diffusion models for image restoration,'' \emph{IEEE
  Transactions on Multimedia}, vol.~26, pp. 7058--7073, 2024.

\bibitem{9941493}
X.~Sheng, J.~Li, B.~Li, L.~Li, D.~Liu, and Y.~Lu, ``Temporal context mining for
  learned video compression,'' \emph{IEEE Transactions on Multimedia}, vol.~25,
  pp. 7311--7322, 2023.

\bibitem{9919402}
J.~Zhu, Q.~Zhang, L.~Fei, R.~Cai, Y.~Xie, B.~Sheng, and X.~Yang, ``Fffn:
  Frame-by-frame feedback fusion network for video super-resolution,''
  \emph{IEEE Transactions on Multimedia}, vol.~25, pp. 6821--6835, 2023.

\bibitem{8723517}
H.~Lin, X.~He, L.~Qing, Q.~Teng, and S.~Yang, ``Improved low-bitrate hevc video
  coding using deep learning based super-resolution and adaptive block
  patching,'' \emph{IEEE Transactions on Multimedia}, vol.~21, no.~12, pp.
  3010--3023, 2019.

\bibitem{10288391}
L.~Tang, Z.~Chen, J.~Huang, and J.~Ma, ``Camf: An interpretable infrared and
  visible image fusion network based on class activation mapping,'' \emph{IEEE
  Transactions on Multimedia}, vol.~26, pp. 4776--4791, 2024.

\bibitem{10239514}
Y.~Xiao, Q.~Yuan, K.~Jiang, X.~Jin, J.~He, L.~Zhang, and C.-W. Lin,
  ``Local-global temporal difference learning for satellite video
  super-resolution,'' \emph{IEEE Transactions on Circuits and Systems for Video
  Technology}, vol.~34, no.~4, pp. 2789--2802, 2024.

\bibitem{9530280}
Y.~Xiao, X.~Su, Q.~Yuan, D.~Liu, H.~Shen, and L.~Zhang, ``Satellite video
  super-resolution via multiscale deformable convolution alignment and temporal
  grouping projection,'' \emph{IEEE Transactions on Geoscience and Remote
  Sensing}, vol.~60, pp. 1--19, 2022.

\bibitem{VC}
M.~Afonso, F.~Zhang, and D.~R. Bull, ``Video compression based on
  spatio-temporal resolution adaptation,'' \emph{IEEE Transactions on Circuits
  and Systems for Video Technology}, vol.~29, no.~1, pp. 275--280, 2019.

\bibitem{face}
H.~Wang and S.~Wang, ``Low-resolution face recognition enhanced by
  high-resolution facial images,'' in \emph{2023 IEEE 17th International
  Conference on Automatic Face and Gesture Recognition (FG)}, 2023, pp. 1--8.

\bibitem{9409729}
C.~Deng, M.~Wang, L.~Liu, Y.~Liu, and Y.~Jiang, ``Extended feature pyramid
  network for small object detection,'' \emph{IEEE Transactions on Multimedia},
  vol.~24, pp. 1968--1979, 2022.

\bibitem{uhd}
S.~Y. Kim, J.~Oh, and M.~Kim, ``Deep sr-itm: Joint learning of super-resolution
  and inverse tone-mapping for 4k uhd hdr applications,'' in \emph{2019
  IEEE/CVF International Conference on Computer Vision (ICCV)}, 2019, pp.
  3116--3125.

\bibitem{9793365}
H.~Wang, W.~Yang, Q.~Liao, and J.~Zhou, ``Bi-rstu: Bidirectional recurrent
  upsampling network for space-time video super-resolution,'' \emph{IEEE
  Transactions on Multimedia}, vol.~25, pp. 4742--4751, 2023.

\bibitem{rbpn}
M.~Haris, G.~Shakhnarovich, and N.~Ukita, ``Recurrent back-projection network
  for video super-resolution,'' in \emph{2019 IEEE/CVF Conference on Computer
  Vision and Pattern Recognition (CVPR)}, 2019, pp. 3892--3901.

\bibitem{tdan}
Y.~Tian, Y.~Zhang, Y.~Fu, and C.~Xu, ``Tdan: Temporally-deformable alignment
  network for video super-resolution,'' in \emph{2020 IEEE/CVF Conference on
  Computer Vision and Pattern Recognition (CVPR)}, 2020, pp. 3357--3366.

\bibitem{edvr}
X.~Wang, K.~C. Chan, K.~Yu, C.~Dong, and C.~C. Loy, ``Edvr: Video restoration
  with enhanced deformable convolutional networks,'' in \emph{2019 IEEE/CVF
  Conference on Computer Vision and Pattern Recognition Workshops (CVPRW)},
  2019, pp. 1954--1963.

\bibitem{basicvsr}
K.~C. Chan, X.~Wang, K.~Yu, C.~Dong, and C.~C. Loy, ``Basicvsr: The search for
  essential components in video super-resolution and beyond,'' in \emph{2021
  IEEE/CVF Conference on Computer Vision and Pattern Recognition (CVPR)}, 2021,
  pp. 4945--4954.

\bibitem{wen2022video}
W.~Wen, W.~Ren, Y.~Shi, Y.~Nie, J.~Zhang, and X.~Cao, ``Video super-resolution
  via a spatio-temporal alignment network,'' \emph{IEEE Transactions on Image
  Processing}, vol.~31, pp. 1761--1773, 2022.

\bibitem{liu2022learning}
C.~Liu, H.~Yang, J.~Fu, and X.~Qian, ``Learning trajectory-aware transformer
  for video super-resolution,'' in \emph{Proceedings of the IEEE/CVF Conference
  on Computer Vision and Pattern Recognition}, 2022, pp. 5687--5696.

\bibitem{qiu2023learning}
Z.~Qiu, H.~Yang, J.~Fu, D.~Liu, C.~Xu, and D.~Fu, ``Learning degradation-robust
  spatiotemporal frequency-transformer for video super-resolution,'' \emph{IEEE
  Transactions on Pattern Analysis and Machine Intelligence}, 2023.

\bibitem{dosovitskiy2015flownet}
A.~Dosovitskiy, P.~Fischer, E.~Ilg, P.~Hausser, C.~Hazirbas, V.~Golkov, P.~Van
  Der~Smagt, D.~Cremers, and T.~Brox, ``Flownet: Learning optical flow with
  convolutional networks,'' in \emph{Proceedings of the IEEE international
  conference on computer vision}, 2015, pp. 2758--2766.

\bibitem{xue2019video}
T.~Xue, B.~Chen, J.~Wu, D.~Wei, and W.~T. Freeman, ``Video enhancement with
  task-oriented flow,'' \emph{International Journal of Computer Vision}, vol.
  127, pp. 1106--1125, 2019.

\bibitem{liu2019deep}
Y.-L. Liu, Y.-T. Liao, Y.-Y. Lin, and Y.-Y. Chuang, ``Deep video frame
  interpolation using cyclic frame generation,'' in \emph{Proceedings of the
  AAAI Conference on Artificial Intelligence}, vol.~33, no.~01, 2019, pp.
  8794--8802.

\bibitem{bao2019depth}
W.~Bao, W.-S. Lai, C.~Ma, X.~Zhang, Z.~Gao, and M.-H. Yang, ``Depth-aware video
  frame interpolation,'' in \emph{Proceedings of the IEEE/CVF Conference on
  Computer Vision and Pattern Recognition}, 2019, pp. 3703--3712.

\bibitem{park2020robust}
M.~Park, H.~G. Kim, S.~Lee, and Y.~M. Ro, ``Robust video frame interpolation
  with exceptional motion map,'' \emph{IEEE Transactions on Circuits and
  Systems for Video Technology}, vol.~31, no.~2, pp. 754--764, 2020.

\bibitem{lee2020adacof}
H.~Lee, T.~Kim, T.-y. Chung, D.~Pak, Y.~Ban, and S.~Lee, ``Adacof: Adaptive
  collaboration of flows for video frame interpolation,'' in \emph{Proceedings
  of the IEEE/CVF Conference on Computer Vision and Pattern Recognition}, 2020,
  pp. 5316--5325.

\bibitem{kong2022progressive}
L.~Kong, J.~Liu, and J.~Yang, ``Progressive motion context refine network for
  efficient video frame interpolation,'' \emph{IEEE Signal Processing Letters},
  vol.~29, pp. 2338--2342, 2022.

\bibitem{zhu2023mfnet}
G.~Zhu, Z.~Qin, Y.~Ding, Y.~Liu, and Z.~Qin, ``Mfnet: Real-time motion focus
  network for video frame interpolation,'' \emph{IEEE Transactions on
  Multimedia}, 2023.

\bibitem{liu2023ttvfi}
C.~Liu, H.~Yang, J.~Fu, and X.~Qian, ``Ttvfi: Learning trajectory-aware
  transformer for video frame interpolation,'' \emph{IEEE Transactions on Image
  Processing}, 2023.

\bibitem{plack2023frame}
M.~Plack, K.~M. Briedis, A.~Djelouah, M.~B. Hullin, M.~Gross, and C.~Schroers,
  ``Frame interpolation transformer and uncertainty guidance,'' in
  \emph{Proceedings of the IEEE/CVF Conference on Computer Vision and Pattern
  Recognition}, 2023, pp. 9811--9821.

\bibitem{shechtman2005space}
E.~Shechtman, Y.~Caspi, and M.~Irani, ``Space-time super-resolution,''
  \emph{IEEE Transactions on Pattern Analysis and Machine Intelligence},
  vol.~27, no.~4, pp. 531--545, 2005.

\bibitem{shechtman2011}
O.~Shahar, A.~Faktor, and M.~Irani, ``Space-time super-resolution from a single
  video,'' in \emph{CVPR 2011}, 2011, pp. 3353--3360.

\bibitem{takeda2010spatiotemporal}
H.~Takeda, P.~Van~Beek, and P.~Milanfar, ``Spatiotemporal video upscaling using
  motion-assisted steering kernel (mask) regression,'' \emph{High-quality
  visual experience: creation, processing and interactivity of high-resolution
  and high-dimensional video signals}, pp. 245--274, 2010.

\bibitem{kang2020deep}
J.~Kang, Y.~Jo, S.~W. Oh, P.~Vajda, and S.~J. Kim, ``Deep space-time video
  upsampling networks,'' in \emph{Computer Vision--ECCV 2020: 16th European
  Conference, Glasgow, UK, August 23--28, 2020, Proceedings, Part X}.\hskip 1em
  plus 0.5em minus 0.4em\relax Springer, 2020, pp. 701--717.

\bibitem{dutta2021efficient}
S.~Dutta, N.~A. Shah, and A.~Mittal, ``Efficient space-time video super
  resolution using low-resolution flow and mask upsampling,'' in
  \emph{Proceedings of the IEEE/CVF Conference on Computer Vision and Pattern
  Recognition}, 2021, pp. 314--323.

\bibitem{xiang2020zooming}
X.~Xiang, Y.~Tian, Y.~Zhang, Y.~Fu, J.~P. Allebach, and C.~Xu, ``Zooming
  slow-mo: Fast and accurate one-stage space-time video super-resolution,'' in
  \emph{Proceedings of the IEEE/CVF conference on computer vision and pattern
  recognition}, 2020, pp. 3370--3379.

\bibitem{haris2020space}
M.~Haris, G.~Shakhnarovich, and N.~Ukita, ``Space-time-aware multi-resolution
  video enhancement,'' in \emph{Proceedings of the IEEE/CVF conference on
  computer vision and pattern recognition}, 2020, pp. 2859--2868.

\bibitem{xu2021temporal}
G.~Xu, J.~Xu, Z.~Li, L.~Wang, X.~Sun, and M.-M. Cheng, ``Temporal modulation
  network for controllable space-time video super-resolution,'' in
  \emph{Proceedings of the IEEE/CVF Conference on Computer Vision and Pattern
  Recognition}, 2021, pp. 6388--6397.

\bibitem{zhang2022cross}
W.~Zhang, M.~Zhou, C.~Ji, X.~Sui, and J.~Bai, ``Cross-frame transformer-based
  spatio-temporal video super-resolution,'' \emph{IEEE Transactions on
  Broadcasting}, vol.~68, no.~2, pp. 359--369, 2022.

\bibitem{10466790}
C.~Fu, H.~Yuan, L.~Shen, R.~Hamzaoui, and H.~Zhang, ``3dattgan: A 3d
  attention-based generative adversarial network for joint space-time video
  super-resolution,'' \emph{IEEE Transactions on Emerging Topics in
  Computational Intelligence}, pp. 1--12, 2024.

\bibitem{XIAO2022102731}
\BIBentryALTinterwordspacing
Y.~Xiao, Q.~Yuan, J.~He, Q.~Zhang, J.~Sun, X.~Su, J.~Wu, and L.~Zhang,
  ``Space-time super-resolution for satellite video: A joint framework based on
  multi-scale spatial-temporal transformer,'' \emph{International Journal of
  Applied Earth Observation and Geoinformation}, vol. 108, p. 102731, 2022.
  [Online]. Available:
  \url{https://www.sciencedirect.com/science/article/pii/S0303243422000575}
\BIBentrySTDinterwordspacing

\bibitem{hu2023store}
M.~Hu, K.~Jiang, Z.~Nie, J.~Zhou, and Z.~Wang, ``Store and fetch immediately:
  Everything is all you need for space-time video super-resolution,'' in
  \emph{Proceedings of the AAAI Conference on Artificial Intelligence},
  vol.~37, no.~1, 2023, pp. 863--871.

\bibitem{hu2023cycmunet+}
M.~Hu, K.~Jiang, Z.~Wang, X.~Bai, and R.~Hu, ``Cycmunet+: Cycle-projected
  mutual learning for spatial-temporal video super-resolution,'' \emph{IEEE
  Transactions on Pattern Analysis and Machine Intelligence}, vol.~45, no.~11,
  pp. 13\,376--13\,392, 2023.

\bibitem{pixelshuffel}
W.~Shi, J.~Caballero, F.~Huszár, J.~Totz, A.~P. Aitken, R.~Bishop,
  D.~Rueckert, and Z.~Wang, ``Real-time single image and video super-resolution
  using an efficient sub-pixel convolutional neural network,'' in \emph{2016
  IEEE Conference on Computer Vision and Pattern Recognition (CVPR)}, 2016, pp.
  1874--1883.

\bibitem{tao2017detail}
X.~Tao, H.~Gao, R.~Liao, J.~Wang, and J.~Jia, ``Detail-revealing deep video
  super-resolution,'' in \emph{Proceedings of the IEEE international conference
  on computer vision}, 2017, pp. 4472--4480.

\bibitem{shi2015convolutional}
X.~Shi, Z.~Chen, H.~Wang, D.-Y. Yeung, W.-K. Wong, and W.-c. Woo,
  ``Convolutional lstm network: A machine learning approach for precipitation
  nowcasting,'' \emph{Advances in neural information processing systems},
  vol.~28, 2015.

\bibitem{lai2017deep}
W.-S. Lai, J.-B. Huang, N.~Ahuja, and M.-H. Yang, ``Deep laplacian pyramid
  networks for fast and accurate super-resolution,'' in \emph{Proceedings of
  the IEEE conference on computer vision and pattern recognition}, 2017, pp.
  624--632.

\bibitem{wang2004image}
Z.~Wang, A.~C. Bovik, H.~R. Sheikh, and E.~P. Simoncelli, ``Image quality
  assessment: from error visibility to structural similarity,'' \emph{IEEE
  transactions on image processing}, vol.~13, no.~4, pp. 600--612, 2004.

\bibitem{haris2018deep}
M.~Haris, G.~Shakhnarovich, and N.~Ukita, ``Deep back-projection networks for
  super-resolution,'' in \emph{Proceedings of the IEEE conference on computer
  vision and pattern recognition}, 2018, pp. 1664--1673.

\bibitem{son2021ntire}
S.~Son, S.~Lee, S.~Nah, R.~Timofte, and K.~M. Lee, ``Ntire 2021 challenge on
  video super-resolution,'' in \emph{Proceedings of the IEEE/CVF Conference on
  Computer Vision and Pattern Recognition}, 2021, pp. 166--181.

\bibitem{soomro2012ucf101}
K.~Soomro, A.~R. Zamir, and M.~Shah, ``Ucf101: A dataset of 101 human actions
  classes from videos in the wild,'' \emph{arXiv preprint arXiv:1212.0402},
  2012.

\bibitem{bao2019memc}
W.~Bao, W.-S. Lai, X.~Zhang, Z.~Gao, and M.-H. Yang, ``Memc-net: Motion
  estimation and motion compensation driven neural network for video
  interpolation and enhancement,'' \emph{IEEE transactions on pattern analysis
  and machine intelligence}, vol.~43, no.~3, pp. 933--948, 2019.

\bibitem{woo2018cbam}
S.~Woo, J.~Park, J.-Y. Lee, and I.~S. Kweon, ``Cbam: Convolutional block
  attention module,'' in \emph{Proceedings of the European conference on
  computer vision (ECCV)}, 2018, pp. 3--19.

\bibitem{qin2021fcanet}
Z.~Qin, P.~Zhang, F.~Wu, and X.~Li, ``Fcanet: Frequency channel attention
  networks,'' in \emph{Proceedings of the IEEE/CVF international conference on
  computer vision}, 2021, pp. 783--792.

\end{thebibliography}

\end{document}